# Quench rate dependence of center formation in Er implanted Si

Mark A. Hughes,[1*] Huan Liu,[2] Yaping Dan[2]

[1]*School of Science, Engineering and Environment, University of Salford, Salford, M5 4WT, UK*
[2]*Global College, Shanghai Jiao Tong University, Shanghai 200240, China*
[*]m.a.hughes@salford.ac.uk

Er implanted Si (Er:Si) is a promising candidate for scalable planar quantum memory (QM) applications. Er has a preference to coordinate with O impurities, and multiple types of Er center are typically formed after a post implant anneal. Float zone Si was implanted with $10^{18}$ cm$^{-3}$ Er and separate samples were annealed using a rapid quench annealing technique at 950 °C for 10 min with quench rates of 5, 23, 46, 93, 185 and 400 °C/s. The evolution of photoluminescence (PL) peaks and their associated Er centers was tracked as a function of quench rate. Across all samples, five distinct Er centers were identified. Two centers, one with mixed Si/O coordination and one with Si-only coordination, exhibited fully resolved crystal-field splitting of the $^4I_{15/2}$ ground state together with 2–3 hot lines from the $^4I_{13/2}$ excited state; fitting of crystal-field parameters for both was consistent with $C_{2v}$ symmetry. The mixed Si/O coordinated center was suppressed at quench rates above 185 °C/s, while the Si-only coordinated center was progressively enhanced with increasing quench rate up to the maximum of 400 °C/s. These results demonstrate that rapid quench annealing has the potential to selectively stabilize a single, Si-coordinated Er center in Er:Si, which is required for QM applications.

## 1. Introduction

For scalability, photonic quantum computers (QCs) should be based on planar waveguide technology. Spontaneous single-photon sources (SPSs) using phenomena such as spontaneous parametric down conversion (SPDC) and spontaneous four wave mixing (SFWM) are favored in many photonic QC architectures because of their high spectral purity and indistinguishability, and ease of integration into planar waveguide devices.
The performance of photonic QCs using spontaneous SPSs can be boosted by using an on-demand QM to synchronize photons from multiple spontaneous SPSs [1]. QMs operating at telecoms wavelengths are favorable since most photonic QCs, and the fiber interconnects for future distributed quantum computing, operate at these wavelengths. However, there is currently no efficient and scalable on-demand QM that can be integrated into planar devices; such a device would revolutionize the design and performance of photonic QCs [2]. Er doped crystals are favorable for telecoms wavelength planar QM [3]. Si is attractive for its pedigree in integrated circuit fabrication, and the precision in placement and concentration of ion implantation makes it the favored technique to introduce impurities into planar devices, making Er implanted Si (Er:Si) a natural candidate to investigate for QM applications.

To fully exploit a scalable planar QM, integration with a suitable spontaneous SPS is essential. Owing to its high nonlinearity, Si-based SFWM devices can be made significantly shorter than those in competing platforms such as SiN, achieving footprints as small as ~0.1 mm$^2$ [4]. Moreover, they can be readily interfaced with Er:Si QMs, providing additional motivation to investigate Er:Si for QM applications.

Currently, compared to Er:Si, the properties of Er doped transparent crystals are favorable for QM applications in terms of coherence: optical $T_2$ (2 ms in Er:Y$_2$SiO$_5$ [5]) and spin $T_2$ (23 ms in Er:CaWO$_4$ [6]), even at relatively high doping concentrations; in terms of center formation, Er:Si suffers from the uncontrollable formation of multiple Er centers, while Er doped transparent crystals generally do not. However, a number of recent advances have been made in Er:Si, including an Er implanted Si waveguides with a 9 kHz homogeneous linewidth [7], a spin $T_2$ of ~10 μs in Er implanted $^{nat}$Si [8] has been extended to ~ms in Er implanted $^{28}$Si [9], spin-resolved excitation of single Er centers with <0.1 GHz spectral diffusion linewidth [10], and we demonstrated the first coupling with superconducting circuits [11].

Crystals with a well-developed photonic technology include Si and, to some extent, LiNbO$_3$, but not the traditional transparent bulk crystal hosts for Er, such as Y$_2$SiO$_5$ or CaWO$_4$. Unlike Si, LiNbO$_3$ is less compatible with current photonic QC architectures. Impurities with a spin tend to exhibit a short spin $T_2$ due to the relatively high nuclear spin of LiNbO$_3$, making spin-wave quantum memory, where optical coherence is transferred to longer-lived spin states via Zeeman or hyperfine-dependent pulses, difficult. Er doped LiNbO$_3$ also has an unusually large inhomogeneous linewidth, reducing the optical depth for a given Er concentration, and LiNbO$_3$ shows poor resistance to Er implantation and annealing [12].

Annealing is a critical process in almost all commercial and research semiconductor technologies involving ion implantation. The standard tool is a rapid thermal annealer (RTA) which uses high power lamps to heat an entire wafer at heating rates of 50–300 °C/s and cooling rates of 10–100 °C/s [13-15]. Usually, only the anneal time and temperature are reported, and the quench rate of the anneal is not given much attention. We recently demonstrated a rapid quench annealing (RQA) technique that has the potential to overcome the issues with Er:Si by producing the isolated Er centers desirable for QM [3]. Our RQA technique uses a modified dilatometer which allows much higher quenching rates (up to 1000 °C/s).

Coimplantation with O has previously been used to obtain photoluminescence (PL) from indirect (above bandgap) excitation of Er:Si; the O coimplantation creates an Er-O defect with a defect state close to the conduction band which facilitates the indirect excitation via excitons and leads to n-type conductivity [16].

A comparison between RQA, with a quench rate of ~1000 °C/s, and RTA on 10$^{19}$ cm$^{-3}$ Er and 10$^{19}$ cm$^{-3}$ O implanted Si has previously been reported; however, for QM applications of Er:Si it is important to avoid the n-type conductivity associated with O implantation since free carriers can lead to waveguide propagation loss and joule heating when electric fields are applied (in certain QM protocols), and lower concentrations are required to avoid detrimental effects on optical and spin $T_2$. Here we use RQA on Er:Si more suitable for QM applications: 10$^{18}$ cm$^{-3}$ Er only implanted Si, and we precisely control quench rate from 400 to 5 °C/s in order to determine the relationship between quench rate and the formation of Er centers, including those that are favorable for QM applications.

## 2. Methods

*2.1. Sample preparation*

Samples were fabricated by implanting Er into a phosphorus-doped, <100>-oriented silicon wafer (500 μm thick), supplied by Topsil, at a temperature of 77 K. The unimplanted wafer had a measured resistivity of 8000 ± 500 Ωcm, corresponding to a phosphorus concentration of 5.5 ± 0.3 × 10$^{11}$ cm$^{-3}$. The O impurity concentration, determined by IR absorption measurements, was ~5×10$^{15}$ cm$^{-3}$. A range of implantation energies was used to achieve a uniform ion distribution to a depth of approximately 1.5 μm, as shown in Supplementary Fig. S1; the implantation doses were selected to produce an Er concentration of ~10$^{18}$ cm$^{-3}$. Isotope-selective implantation was employed using only the spin-zero isotope, $^{166}$Er.

Annealing was performed using a rapid thermal annealer (RTA) with a peak cooling rate of 50 to 100 °C/s at a temperature of 750 °C, for 2 minutes; this sample is referred to as

RTA750. Rapid quench annealing (RQA) was carried out using a modified dilatometer (DIL 805A, TA Instruments), in which samples (maximum width: 3 mm) were annealed by induction heating under a vacuum of $5 \times 10^{-4}$ mbar. Rapid cooling was achieved by flushing the chamber with high-purity helium gas (99.999%) pre-cooled in liquid nitrogen (77 K). A K-type thermocouple was used to monitor the sample temperature, which was controlled by adjusting the helium flow rate; RQA anneals were conducted at 950 °C for 10 minutes. The samples were cooled by a two-stage process, the first stage from 950 °C to 25 °C was with a variable controlled cooling time. The second stage is from 25 °C to -100 °C, at a rate of 5 °C/s, is the same for all the samples. The reported cooling rate refers to the initial peak cooling rate, which slows during the initial cooling phase. The RQA annealed samples with initial cooling rates of 400, 185, 93, 46 23 and 5 °C/s are referred to as RQ400, RQ185, RQ93, RQ46, RQ23 and RQ5, respectively; the measured temperature decay profiles for these samples are given in Supplementary Fig. S2. In addition, we used a sample with $10^{19}$ cm$^{-3}$ Er and $10^{19}$ cm$^{-3}$ O implanted Si annealed with RQA with a quench rate of 1000 °C/s which is referred to as DC950. These sample details are summarized in Table 1.

Table 1 Sample details including implanted Er and O concentration and anneal temperature time and quench rate.

| Sample name | Er concentration (cm$^{-3}$) | O concentration (cm$^{-3}$) | Anneal temp (°C) | Anneal time (min) | Quench rate °C/s |
| --- | --- | --- | --- | --- | --- |
| RQ400 | $10^{18}$ | - | 950 | 10 | 400 |
| RQ185 | $10^{18}$ | - | 950 | 10 | 185 |
| RQ93 | $10^{18}$ | - | 950 | 10 | 93 |
| RQ46 | $10^{18}$ | - | 950 | 10 | 46 |
| RQ23 | $10^{18}$ | - | 950 | 10 | 23 |
| RQ5 | $10^{18}$ | - | 950 | 10 | 5 |
| RTA750 | $10^{18}$ | - | 750 | 2 | 50-100 |
| DC950 | $10^{19}$ | $10^{19}$ | 950 | 10 | 1000 |

*2.2. Optical measurements*

PL spectra were obtained by placing the samples in a closed-cycle helium cryostat with a base temperature of 3.5 K. Fluorescence was excited using a 520 nm, 30 mW laser diode, powered by a laser diode driver that was electronically amplitude modulated by a function generator to give an overall fall time of ~1 µs. The PL was dispersed in a Bentham TMc300 monochromator with a 600 line/mm grating blazed at 1600 nm giving a linear dispersion of 5.4 nm/mm and detected with an IR sensitive photomultiplier tube (PMT); the signal was recovered with a SR830 lock-in amplifier. All spectra were corrected for the system response.
    Transient fluorescence measurements were taken with the same system used for PL measurements, except the transient signal was captured with a 500 MHz oscilloscope, the overall system response time was ~1 µs.

*2.3. Crystal field analysis*

Electron interactions in the *f*-orbitals of rare-earth (RE) ions produce relatively local environment (crystal field) independent energy manifolds, unique to each ion, spanning ~$10^4$ cm$^{-1}$; the crystal field does however cause splitting of these manifolds that spans ~$10^2$ cm$^{-1}$ and this splitting is highly sensitive to variations in the crystal field. We use this phenomenon to analyze the Er centers from their PL spectra. Analytically, the Hamiltonian (*H*) of a RE can be described as:

$$H = H_F + H_{CF} \tag{1}$$

Here, $H_F$ describes the interactions within a free RE ion, with parameter sets that are specific to each RE ion and change little between hosts. In our fitting procedure, $H_F$ was not included since we have limited observables and it does not depend significantly on the crystal field. $H_{CF}$ describes the crystal field surrounding a RE ion, and is expressed as the linear combination of spherical tensors, $C_q^{(k)}$, of various ranks and crystal field parameters (CFPs), $B_q^k$, as shown in Eq. 2:

$$H_{CF} = \sum_{k,q} B_q^k C_q^{(k)} \tag{2}$$

Each crystallographic point group has a characteristic set of CFPs [17], and the construction of $H_{CF}$ is described in detail elsewhere [18]. Every $^{2S+1}L_J$ manifold has its own $H_{CF}$ with the crystal field splitting determined by its eigenvalues. To extract CFPs from experimentally measured splitting, we applied a least-squares algorithm that adjusted the $B_q^k$ to minimize the difference between the observed splitting and the eigenvalues of $H_{CF}$.

### 3. Results and discussion

In Er:Si, Er will preferentially coordinate with O over Si when using standard RTA anneals [3]. For QM applications of Er:Si, we are interested in eliminating as many O-containing Er centers as possible. It could be possible for O to diffuse from the bulk silicon toward the Er implanted at the surface during annealing, but this would be limited to a depth of ~10 μm [19-21]. Given the background O concentration of our FZ wafer is $5\times10^{15}$ cm$^{-3}$, the available amount of O diffused from the bulk is at most ~$10^{12}$ cm$^{-2}$ compared with the implanted Er dose of $2.6\times10^{14}$ cm$^{-2}$. We performed a HF etch prior to implant; however, the native oxide will self-limit its growth in a matter of hours at a thickness of ~2 nm [22, 23]. If we assume a 2 nm native oxide prior to implant, there is, in principle, ~$10^{16}$ cm$^{-2}$ O available on the surface. We performed a SRIM simulation of O recoil implantation from a 2 nm native oxide layer into Si. The depth of the recoil implant was ~5 nm, see Supplementary Fig. S3, and the weighted average sputter yield of our Er chain implant was 6.2 O atoms/Er ion. The simulated Er implant profile in Supplementary Fig. S1 shows that the Er is at a depth of ~40 to 1500 nm, so there is not any overlap with Er implants and O recoil implants, but its proximity to the implanted Er makes diffusion from the native oxide during annealing the most likely source of O.

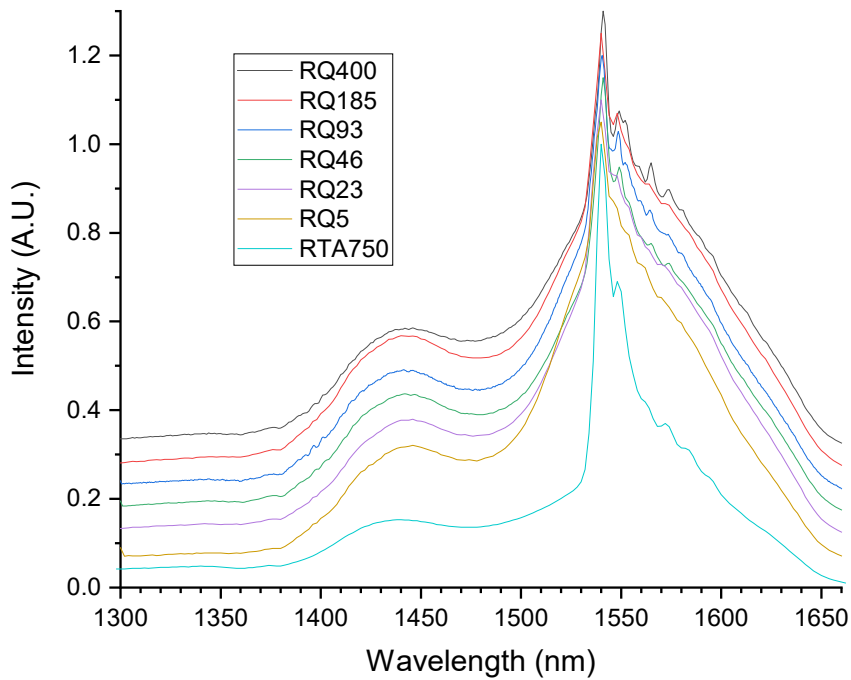

Fig. 1 PL at 3.5 K from 520 nm excitation of $10^{18}$ $cm^{-3}$ Er:Si annealed at 950 °C for 10 min at various quench rates and by standard RTA at 750 °C for 2 min. The resolution was 2.7 nm. The spectra are offset for clarity.

Fig. 1 shows low resolution (2.7 nm) PL from RQ400 to RQ5 samples and RTA750. We are able to obtain relatively strong Er PL peaks from the RQ400 to RQ5 samples, but the Er PL sits on two broad, approximately Gaussian, background bands centered at 1560 nm and 1440 nm. Gaussian fitting in Supplementary Fig. S4 shows the background Gaussians have centers at 6425 $cm^{-1}$, and 6937 $cm^{-1}$, and widths of 350 $cm^{-1}$ and 303 $cm^{-1}$, respectively. The broad background in the RQ400 to RQ5 samples is relatively unaffected by quench rate, however, the background is heavily suppressed in the RTA750 sample, which shows that anneal temperature, or possibly time, affects the background, but quench rate does not.

  PL from above bandgap excitation of Er implanted Si is usually obtained with coimplantation of O, which usually results in well separated, narrow Er peaks with no, or very little, background PL [24-27]. This indicates that our broad background is caused by defects that don't include O.

  In the literature, the requirement for O coimplantation is typically described as the creation of a defect state that allows indirect excitation of Er [16], which indicates that the PL we observe is enabled by an alternative defect state to that with O coimplantation, and that this defect state could be related to the broad background we observe. Supplementary Fig. S5 shows the PL spectrum of RQ400 taken at modulation frequencies of 122 and 1406 Hz. The 122 Hz modulation frequency was slow enough that there was no suppression of any signal, whereas at 1406 Hz, longer PL lifetimes are suppressed compared to shorter lifetimes, and the signal at the Er PL peak at1540 nm was approximately halved. The ratio of these two spectra (122 Hz/1406 Hz) is flat over the two broad bands, showing that both have the same PL wavelength independent lifetime; there are positive peaks corresponding to the narrow Er PL, showing that the Er PL has a longer lifetime than the broad background.

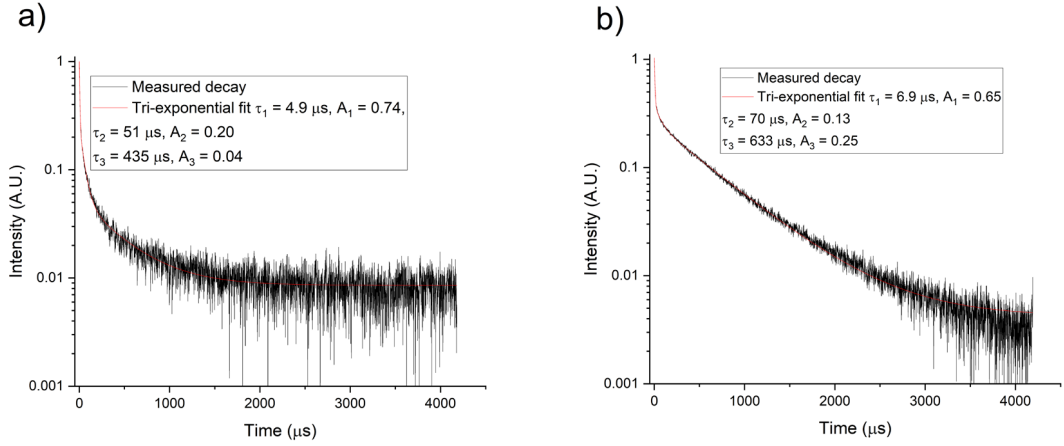

Fig. 2 PL decay profile, fitted with a tri-exponential, measured at a) 1440 nm and b) 1540 nm. The resolution was 10 nm. Temperature was 3.5 K.

Fig. 2 (a) and (b) show the PL decay profile of RQ400 at 3.5 K, measured at wavelengths of 1440 nm and 1540 nm, respectively, which coincide with the peak of the first broad Gaussian background band and at the peak of the Er PL, respectively. The monochromator slits were widened to increase the signal, giving a resolution of ~10 nm. A 10 kΩ shunt resistor was used at the input of the oscilloscope, giving a system response time of ~1 µs. The decays were fitted with increasing numbers of exponential decays until a fit with $R^2>0.99$ was achieved [28], which occurred with a tri-exponential, see Eq. 1.

$$I(t) = A_1 \exp\left(-\frac{t}{\tau_1}\right) + A_2 \exp\left(-\frac{t}{\tau_2}\right) + A_3 \exp\left(-\frac{t}{\tau_3}\right) + y_0, \quad (1)$$

Where $\tau_{1,2,3}$ are the characteristic lifetimes, $A_{1,2,3}$ are their respective coefficients and $y_0$ is the offset. Fig. 2 (a) contains two short lifetime components with lifetimes of 5 and 50 µs, the A coefficients, which give the relative initial intensity of the decay, were 0.75 and 0.2, respectively, there is also a relatively insignificant long lifetime component. Fig. 2 b) contains two short lifetime components with lifetimes of 7 and 70 µs, these two lifetimes, and their intensities relative to each other are approximately the same as in Fig. 2 (a). There is also a long lifetime component of 633 µs, since the Er peaks have a longer lifetime than the background, this must be from the Er peaks. The intensity (the A-coefficient) of this long lifetime decay is 0.25, which is also consistent with the relative intensity of the Er peaks in PL spectrum in Fig. 1. Since there must be an indirect excitation mechanism to enable us to observe Er PL with above bandgap excitation, the broad bands are attributed to PL from the defect states involved in indirect Er excitation; the lifetimes of these broad bands are longer than the ~µs lifetime reported for defect states involved in indirect excitation of Er/O implanted Si [16]. The Er lifetime is shorter than that observed in Er doped transparent crystals (1-5 ms), which we attribute to competing fast radiative and nonradiative decays from the defect states associated with the broad background PL.

Various PL lines associated with dislocation defects have been identified in irradiated or thermally shocked silicon: $D_1$ (1530 nm), $D_2$ (1420 nm), $D_3$ (1340 nm), and $D_4$ (1240 nm) [29]. However, their observation does not depend strongly on the method used to introduce them [30]. The $D_1$ line originates from the core structure of a shuffle Lomer dislocation, which is composed of tightly bonded five- and seven-membered atomic rings. In contrast, the $D_2$ line arises from metastable interstitial defects located on the (001), (111), and (113) planes in the vicinity of various dislocation cores [30]. Our broad background peaks at 1560 and 1440 nm are a good match to those reported for $D_1$ and $D_2$ lines, respectively. In Si implanted Si, PL has been reported for various Si doses and annealing temperatures [31], which showed that the intensity of $D_1$ and $D_2$ lines increased with Si dose and annealing temperature; for example, with a dose of $3\times10^{15}$ cm$^{-2}$ Si, both $D_1$ and $D_2$ were observed with an annealing temperature of 900 °C, but not 700 °C [31], which is consistent with our observation of the suppression of the broad background with a 750 °C anneal. The time-resolved PL of $D_1$ and $D_2$ defects at 6 K have short and long lifetime components of 4.7 to 10 µs and 40 to 1900 µs, respectively [32], which is consistent with our lifetime measurements of the broad background. These dislocations, although unwanted for QM applications, may allow us to observe PL through simple indirect excitation, rather than through direct excitation that requires waveguide fabrication.

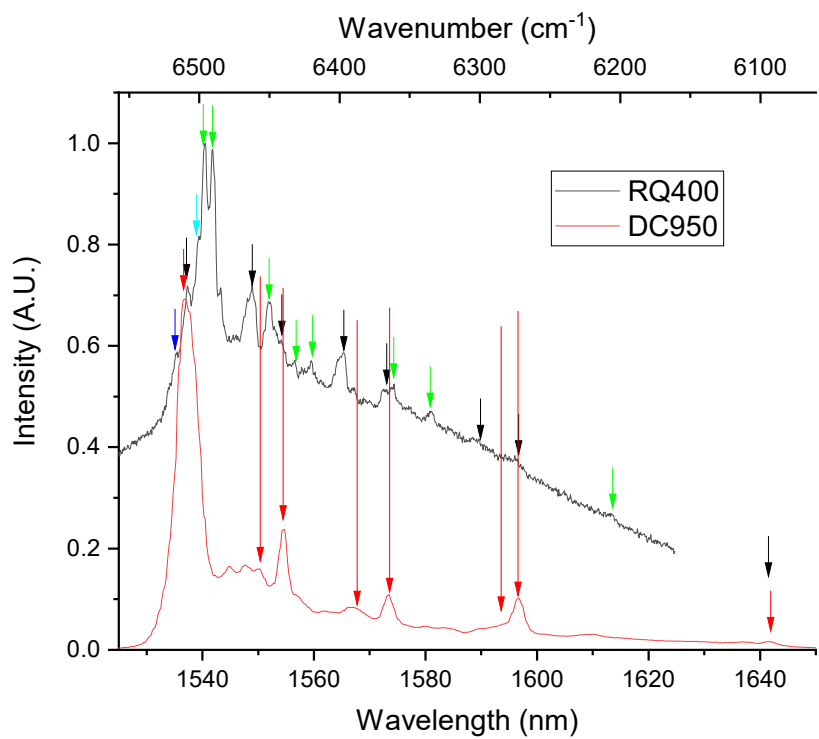

Fig. 3 High resolution (0.6 nm) PL spectra of RQ400 and DC950. Red arrows show the peaks for the Er-C center that has previously been identified in the DC950 sample [3], black arrows show the peaks in RQ400 that correspond to the RQ-C center, green arrows correspond to the RQ-E center. Dark blue and light blue arrows correspond to the RQ-A and RQ-D centers, respectively.

As shown in Fig. 3, the PL spectrum of RQ400 is complex, with many more peaks than the eight expected for a single Er center [3]. There are four peaks at 1535.4 nm, 1537.3 nm, 1539.3nm , and 1540.5 nm, which are in the region we would typically expect to find the strongest transition of a single Er center – the transition from the lowest crystal field level of the excided state manifold to the lowest crystal field level of the ground state manifold, or zero phonon line (ZPL). In previous work, the peaks of an Er center, called Er-C, with $C_{2v}$ symmetry were identified in the DC950 sample [3]. In Fig. 3 the peaks belonging to Er-C are identified with red arrows

      Non-ZPL peaks (transitions from the lowest crystal field level of the excided state manifold to higher crystal field levels of the ground state manifold) are strongly affected by

CFPs which can vary significantly between centers. The ZPL peak effectively represents the energy separation of the ground an excited state manifold. Differences in the ZPL between different Er centers are caused by both CFPs, which don't affect it much, and free ion parameters, which vary very little between host materials. So a change to the local environment of an Er center will have a relatively small effect on the ZPL peak, and a much larger effect on non-ZPL peaks. Therefore, the wavelength of the ZPL is a relatively salient property of a particular Er center. The ZPL of Er-C matches the peak at 1537.3 nm in RQ400 very well; there are other peaks in RQ400 close to those in Er-C that we have identified with black arrows. The signal for RQ400 was weaker than for DC950, and long wavelength peaks become progressively weaker, even in the low resolution spectrum with higher signal in Fig. 1, the long wavelength peak at 1642 nm couldn't be resolved, so we assume the same peak position as Er-C. The strongest peak in RQ400 is at 1540.5 nm, we assume this is the ZPL of another Er center, and the peaks not assigned to the 1537.3 nm ZPL center are assigned to this center and identified with green arrows. The peaks at 1535.4 nm and 1539.3 nm are assumed to be ZPLs of two more centers with no other resolvable peaks.

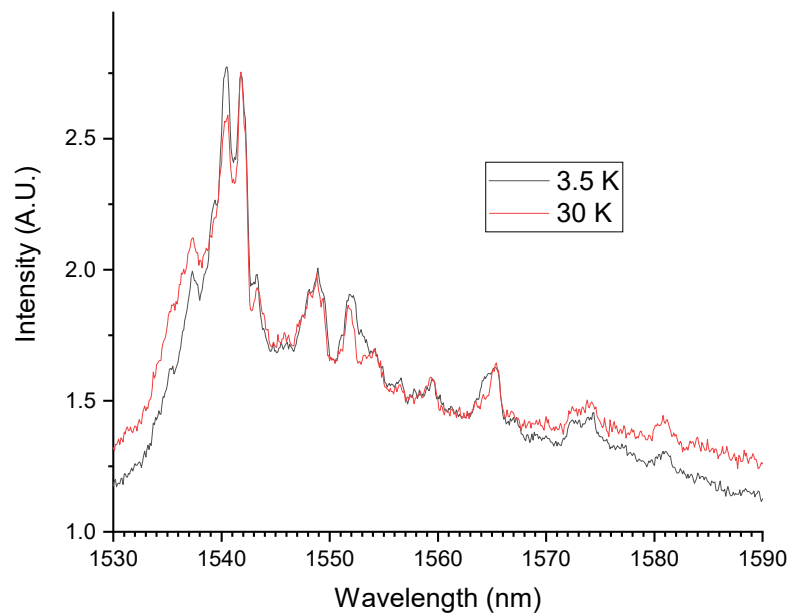

Fig. 4 PL spectra of RQ400 at 3.5 K and 30 K, normalized to the 1541.8 nm peak. Resolution was 0.6 nm.

Fig. 4 shows the PL spectra of RQ400 at 3.5 K and 30 K. At increasing temperatures, we would expect thermal effects to increase the intensity of non-ZPL peaks relative to the ZPL peak. Comparing the relative intensities of the peaks at the two temperatures was complicated by a change in the background PL; however, by normalizing at the 1541.8 nm peak, all the proposed non-ZPL peaks (1541.8 nm and longer) have approximately the same relative intensity, and all the proposed ZPL peaks 1535.4 nm to 1540.5 nm, have decreased in intensity (their height above the background) indicating they are all ZPLs.

In order to distinguish Er centers based on the principle that different Er centers can have different PL lifetimes, we measured the lifetime of RQ400 at various narrow Er PL peaks. In order to achieve spectral resolution of the peaks we reduced the monochromator slit width to give a resolution of 0.6 nm. We sacrificed the ability to resolve the short lifetime components, which isn't relevant for this measurement, for increased signal by increasing the oscilloscope shunt resistance to 100 kΩ. Supplementary Fig. S6 shows the fitted long lifetime component for various Er PL peaks. It confirms that the 1535.4 nm peak can be identified as the ZPL for a separate center; however, all other peaks cannot be distinguished in terms of lifetime.

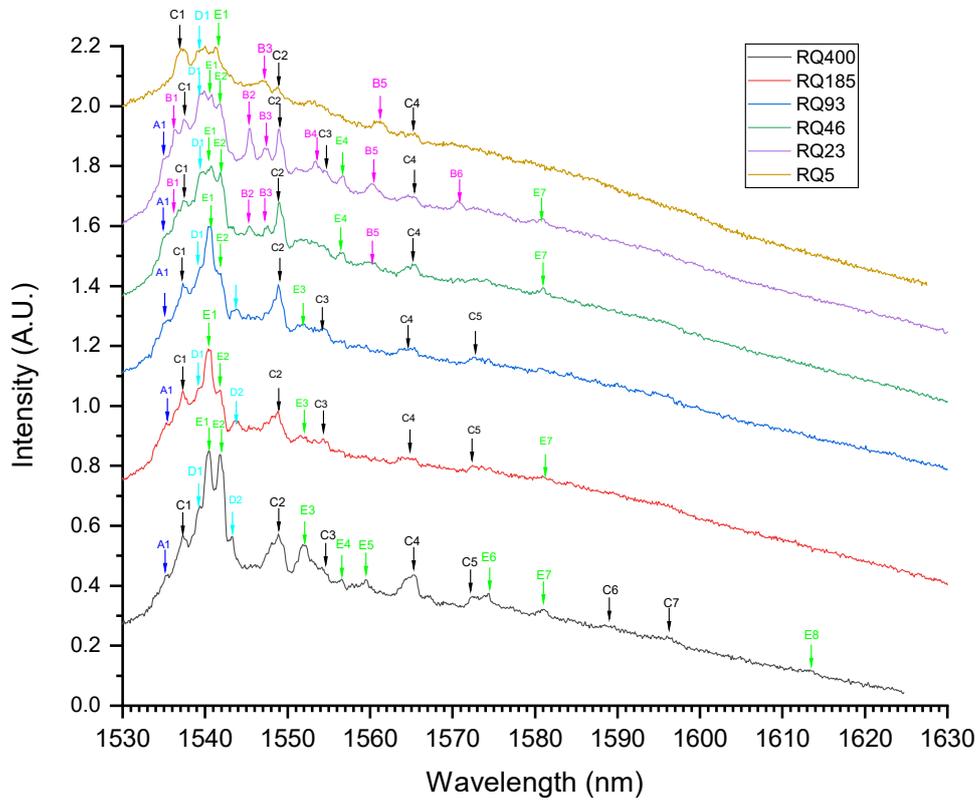

Fig. 5 PL of RQ400 to 5 samples at 3.5 K. The resolution was 0.6 nm. The peaks for each center in each sample are identified with indexed colored arrows. RQ-A, blue; RQ-B, purple; RQ-C, black; RQ-D, light blue; RQ-E, green.

Fig. 5 shows high resolution (0.6 nm) PL spectra of all the rapid quench samples with various quench rates, offset from each other for clarity. We identify five distinct Er centers that are named RQ-A to E in the order of their ZPL wavelength, see Table 2; note that the 1536.4 nm ZPL peak is first visible in RQ93, all others can be resolved in RQ400.

Table 2 Er center PL wavelengths (nm).

| Manifold | Peak | Er Center | | | | |
| --- | --- | --- | --- | --- | --- | --- |
| | | RQ-A | RQ-B | RQ-C | RQ-D | RQ-E |
| $^4I_{15/2}$ | 1 | 1535.4 | 1536.4 | 1537.3 | 1539.3 | 1540.5 |
| | 2 | | 1545.3 | 1549.0 | 1543.5 | 1541.8 |
| | 3 | | 1547.4 | 1554.3 | | 1552.1 |
| | 4 | | 1553.3 | 1565.2 | | 1556.6 |
| | 5 | | 1560.3 | 1574.2 | | 1559.5 |
| | 6 | | 1570.7 | 1589.1 | | 1574.3 |
| | 7 | | | 1596.2 | | 1581.0 |
| | 8 | | | 1642.2 | | 1613.4 |
| $^4I_{13/2}$ | 1 | | | 1519.4 | | 1531.1 |
| | 2 | | | 1511.0 | | 1527.6 |
| | 3 | | | | | 1524.2 |

The wavelength of the peaks does not change significantly when tracked across the quench rates, but relative intensities vary significantly and peaks can appear or disappear, which implies that the Er centers in the rapid quench samples are all the same, but their relative abundance varies significantly. There are some significant changes that occur going from RQ400 to RQ185: the relative intensity of the 1537.3 nm ZPL peak (RQ-C) compared to the 1540.5 nm ZPL peak (RQ-E) increases significantly. This can be used to confirm the peak attribution to RQ-E since the non-ZPL peaks that get weaker going from RQ400 to RQ185 should be attributed to RQ-E. As seen in Fig. 5, this is confirmed, which confirms our peak attribution to RQ-C and RQ-E. This leaves only one unattributed peak at 1543.5 nm, this does not decrease significantly between RQ400 and RQ185 and it isn't close to previous Er-C peaks, so we attribute it to the 1539.3 nm ZPL (RQ-D), and it is labelled D2. There is relatively little change going from RQ185 to RQ93; however, at RQ46 new peaks at 1545.3 nm and 1547.4 nm, along with a peak in the expected ZPL region at 1536.4 nm are observed. This is attributed to the appearance of a new Er center (RQ-B) with a ZPL of 1536.4 nm. In RQ23, these peaks get stronger and other new peaks appear which are also attributed to RQ-B, see Table 2. Finally, in RQ5, many peaks are unresolved and the RQ-C,D and E ZPL peaks have approximately equal intensity.

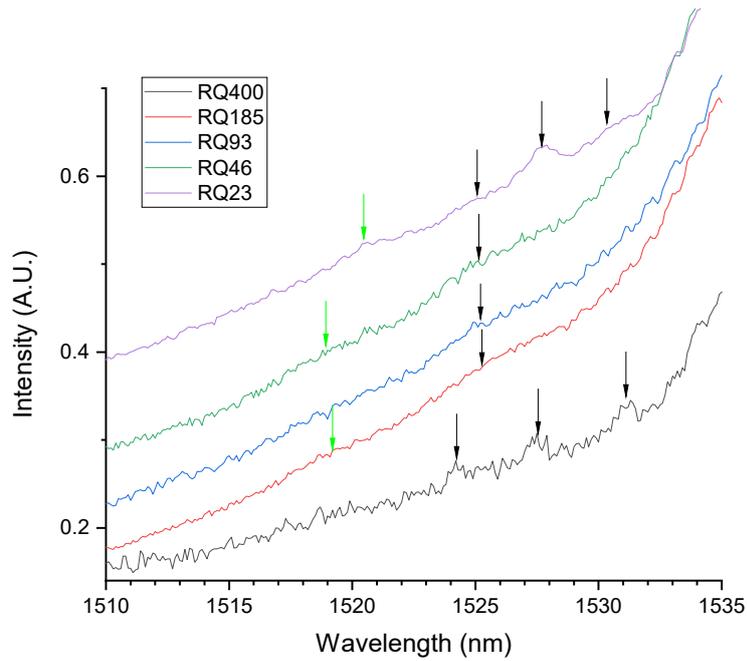

Fig. 6 PL of RQ400 to 23 samples at 70 K. The resolution was 1.2 nm. Black and green arrows indicate hot lines attributed to the RQ-E and RQ-C centers, respectively. No hot lines could be resolved in RQ5.

To identify hot-lines attributed to transitions from higher crystal field levels in the excited state manifold, we measured PL at 70 K, as shown in Fig. 6. To compensate for the decrease in signal, the resolution was increased to 1.2 nm. Clear hot lines for the Er-C center at 1511 and 1520 nm have previously been reported [3], which, as shown in Fig. 3, is very similar to the RQ-C center. A 1511 nm peak cannot be resolved in any sample. A 1520 nm peak is not resolved in RQ400, but is in RQ185, as would be expected for a relative increase in the RQ-C center in RQ185. The three other hot-line peaks at 1531.1 nm, 1527.6 nm and 1524.2 nm are attributed to RQ-E, the only other well resolved center.

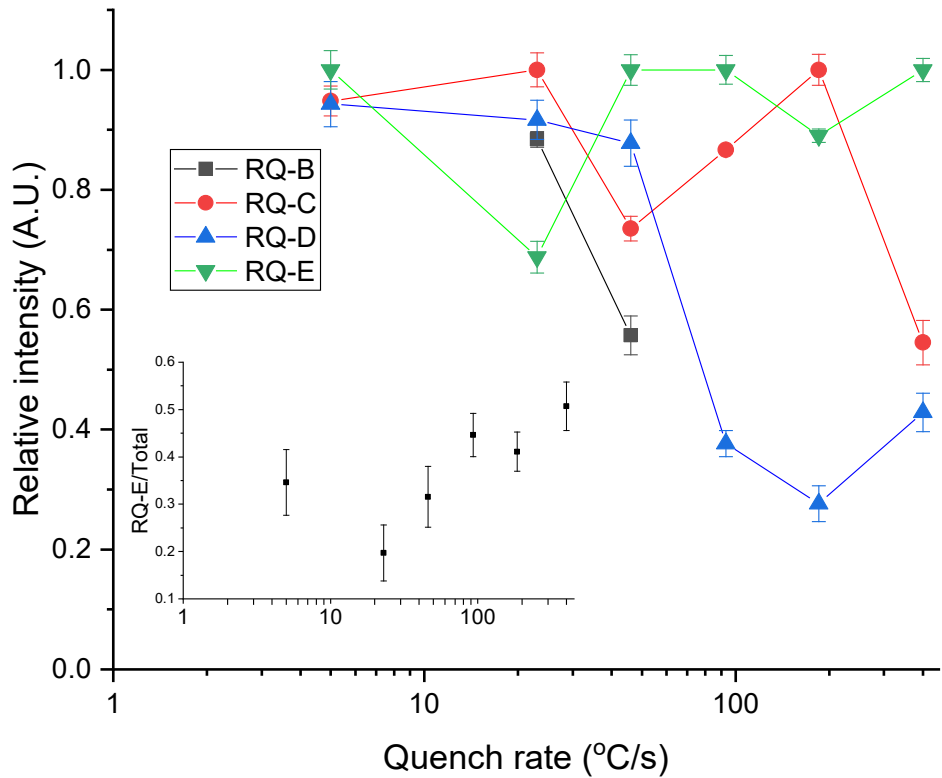

Fig. 7 Quench rate dependence of normalized ZPL intensity at 3.5 K for the RQ-B,C,D and E centers. Inset shows the ratio of the RQ-E ZPL intensity to the total intensities of all visible ZPLs.

The RQ-C center has a similar crystal field splitting to Er-C, which it was previously argued had a mixed Si and O coordination [3], which suggests that RQ-C also has mixed Si and O coordination. XAFSS measurements of Er only implanted FZ Si show only Si coordinated Er [33], within the sensitivity of the measurement. If RQ-C also has a mixed Si and O coordination, given the limited O content, it is likely that RQ-E has Si only coordination.

      The intensity of the ZPL of a particular Er center should scale linearly with its abundance. Determining how the relative intensities of ZPLs attributed to each center vary with quench rate requires distinguishing the Er PL peak from the background. We found deconvolution into Gaussians to be an unreliable method, especially because the background under the Er peaks appears to be more complex than a single Gaussian. Instead, we used high resolution lifetime measurement of the ZPL peaks of the RQ-B, C, D and E centers at 1536.4 nm, 1537.3 nm, 1539.3 nm and 1540.5 nm, respectively. By fitting a double exponential, the A-coefficient of the long lifetime component attributed to Er gives the relative intensity of the Er peak compared to the short lifetime broad background; the A coefficients for the four wavelengths are then normalized to the largest A-coefficient to eliminate the effect of the broad background. These normalized A coefficients are shown in Fig. 7, which highlights the shift in relative Er center abundance caused by the quench rate. Overall, the RQ-B, RQ-C and RQ-D centers tend to be suppressed when the quench rate is increased to 46, 400 and 93 °C/s, respectively. The highest quench rate, 400 °C/s, favors the RQ-E center. The inset of Fig. 7 shows the ratio of the RQ-E A-coefficient to the A-coefficients of all the centers (i.e. what fraction the RQ-E center is of all the Er centers present) as a function of quench rate. There is a steady increase in the fraction of RQ-E with

increasing quench rate, indicating we can increase the purity of the RQ-E center by increasing the quench rate further.

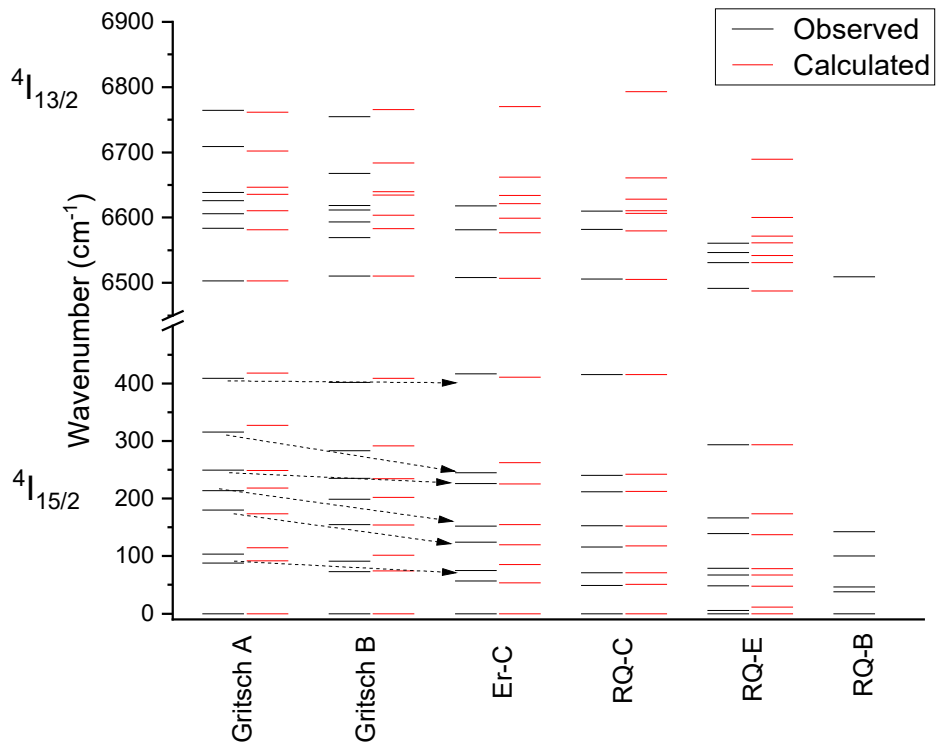

Fig. 8 Observed and calculated energy levels of the Gritsch A, Gritsch B, Er-C, RQ-C and RQ-E centers; the fits have a root mean square deviation (RMSD) of 7.2, 12.1, 8.2, 1.8 and 3.3, respectively. Arrows indicate the trends in energy levels

Gritsch *et al* reported six energy levels of the ground state manifold and all seven energy levels of the excited state manifold, from PL and excitation measurements of Er implanted Si waveguides, for two well defined Er centers, Gritsch A and B [7]; these excited state energy levels are a good match to our indirect measurement of the excited state energy levels by Optically Modulated Magnetic Resonance (OMMR) [3, 26]. Related Er centers can be arranged in a sequence that reflects a monotonic variation in their local environment [3]. For instance, a systematic change in the local environment, such as a gradual alteration in the primitive cell dimensions or a shift in coordination from Si to O, would be expected to produce a corresponding monotonic trend in the fitted CFPs [18]. In previous work, it was shown how these centers, along with the Er-C center could be arranged in a sequence of consistently evolving energy levels and their associated CFPs [3]; a method of descending symmetries was used to show that the symmetry of these centers was orthorhombic $C_{2v}$, or lower [3]; it was proposed that the cause of the monotonic shift was increased Si coordination towards Gritsch A [3]. Subsequently, the final two energy levels of the ground state manifold for the Gritsch A and B centers have been reported [34]. Here, we performed a $C_{2v}$ fit on the updated Gritsch A and B centers and give the previous fit for Er-C in Fig. 8, along with $C_{2v}$ fits to the RQ-C and RQ-E centers, we also show the energy levels of the incomplete RQ-B center, which has insufficient observables for a $C_{2v}$ fit.

      The Er-C and RQ-C centers are very similar in terms of their peak positions, so they should be closely related centers, but the $C_{2v}$ fit for RQ-C is significantly better than for Er-C

(RMSD = 1.8 vs 8.2). The excellent fit of RQ-C indicates correct peak and symmetry assignments. The good, but not excellent, fits for the other centers indicates the symmetry and peak assignments are mostly correct, but leaves room for them to have a symmetry closely related to $C_{2v}$ and/or a small number of incorrectly assigned peaks.

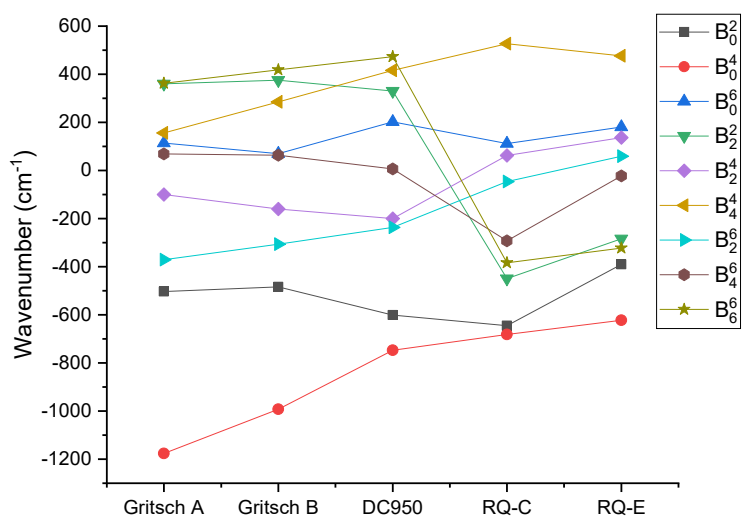

Fig. 9 $C_{2v}$ Orthorhombic CFPs from fits shown in Fig. 8.

Fig. 9 shows how $C_{2v}$ CFPs vary across the centers. The updated CFPs for Gritsch A and B don't change significantly, and the trend in CFPs on the sequence Gritsch A, B, Er-C remains as previously identified [3]. The values of the CFPS are given in Supplementary Table S1.

CFPs are determined by the positions and strengths of electrostatic ligand fields surrounding the RE center; however, a particular set of CFPs doesn't necessarily correspond to a unique local structure. Calculation of CFPs from electrostatic models can help to determine certain aspects of the local structure, such as whether the center in interstitial or substitutional [18], but it is required to already have a general idea of what and how many atoms are coordinating the center and what their distances and positions are. Here we only have some idea of the symmetry. Nevertheless, it is possible to get a qualitative understanding of local structure from comparisons of CFPs.

In $C_{2v}$ symmetry, the $B_0^2$ parameter is sensitive to the overall elongation or compression of the coordination polyhedron along the principal $C_2$ axis (conventionally the z-axis). A positive $B_0^2$ can indicate an elongation of the coordinating ligands along the z-axis, while a negative value suggests a compression [35].

Since RQ-C has a significantly better fit to $C_{2v}$ CFPs than Er-C the sign flip could be caused by Er-C not being $C_{2v}$ symmetry which could give a fitting artefact due to an incomplete set of CFPs.

Comparing RQ-C to RQ-E, all CFPs except $B_0^6$ and $B_2^4$ decrease in magnitude with $B_0^2$ and $B_4^6$ decreasing significantly in RQ-E. Si coordination is expected to result in smaller CFPs than O. CFPs depend strongly on the ligand's charge, electronegativity, and distance from the RE ion. $O^{2-}$ produces a stronger point-charge-like field due to its high ionic character [36, 37] Si-RE bonding is more covalent, with more diffuse charge distribution which gives a weaker multipole interaction with the 4$f$ shell [38]. RQ-E could have a swap of one or few O for Si along main symmetry axis resulting in smaller $B_0^2$. The longer wavelength ZPL of RQ-E also indicates a weaker crystal field that would be expected of Si coordination. This indicates that higher quench rates favor our proposed all (or high) Si coordinated RQ-E center.

## 4. Conclusions

The quench rate of post implant anneals is often given little attention. Er:Si can be viewed as an exemplar system to investigate the effect of quench rate, since many different centers can be formed by different implant and anneal conditions. Er:Si is also promising for QM applications, but requires a single Si only coordinated Er center to be formed.

We obtained high resolution PL from $10^{18}$ cm$^{-3}$ Er only implanted Si annealed with quench rates of 5 to 400 °C/s. Broad bands centered at 1560 nm and 1440 nm were observed and attributed dislocation defects $D_1$ and $D_2$, respectively; these defects could be involved in the indirect excitation mechanism of Er in our samples. Lower temperature anneals can suppress these dislocation defects.

Based on comparisons to previous work, together with temperature dependence, quench-rate dependence, and lifetime measurements, five distinct centers have been identified and designated RQ-A through RQ-E, ordered by their ZPL wavelengths at 1535.4, 1536.4, 1537.3, 1539.3, and 1540.5 nm, respectively. Among these, RQ-C and RQ-E exhibit fully resolved ground-state splittings along with 2-3 hot lines, which can be accurately modelled using a set of CFPs consistent with $C_{2v}$ symmetry, the most likely symmetry assignment for Er centers in Si. RQ-C is attributed to mixed Si and O coordination, based on its close correspondence to previously reported centers, whereas RQ-E is assigned to Si-only coordination, inferred from its small CFP magnitudes and the low oxygen content of our samples. Importantly, RQ-E represents a promising candidate Er center for QM applications in Er:Si. The centers RQ-B, RQ-C, and RQ-D are progressively suppressed at quench rates of approximately 46, 400, and 93 °C/s, respectively, while RQ-E is continuously enhanced up to the maximum quench rate investigated (400 °C/s). This behavior highlights the utility of rapid quench annealing as a means of selectively stabilizing a single, Si-coordinated Er center—an essential requirement for scalable QM implementations in Er:Si.

These results demonstrate the pronounced influence of anneal quench rate on the formation of optically active centers in implanted materials, a principle that may be extended to a wide range of materials systems. In the present samples, the only plausible source of oxygen is diffusion from the native oxide during annealing. To suppress the formation of unwanted O-coordinated centers, it may be necessary to develop a technique enabling rapid transfer to the annealer immediately following the HF dip. However, for practical QM devices, silicon-on-insulator (SOI) substrates would be employed for waveguide fabrication. In SOI, a residual oxygen background is likely unavoidable due to O diffusion into the thin silicon device layer during both the thermal oxidation step of fabrication and the post-implant anneal. This underscores the importance of rapid quench annealing as a route to selectively stabilize a single, Si-coordinated Er center in Er:Si.

**Disclosures**

The authors declare no conflicts of interest.

**Data Availability**

The datasets generated during the current study are available in the Mendeley Data repository https://data.mendeley.com/drafts/kpk88dphhn

**Acknowledgements**

This work was supported by the UK EPSRC grant EP/R011885/1.

**Supplementary Information**
**Quench rate dependence of center formation in Er implanted Si**

Mark A. Hughes,[1*] Huan Liu,[2] Yaping Dan[2]
[1]*School of Science, Engineering and Environment, University of Salford, Salford, M5 4WT, UK*
[2]*Global College, Shanghai Jiao Tong University, Shanghai 200240, China*
[*]m.a.hughes@salford.ac.uk

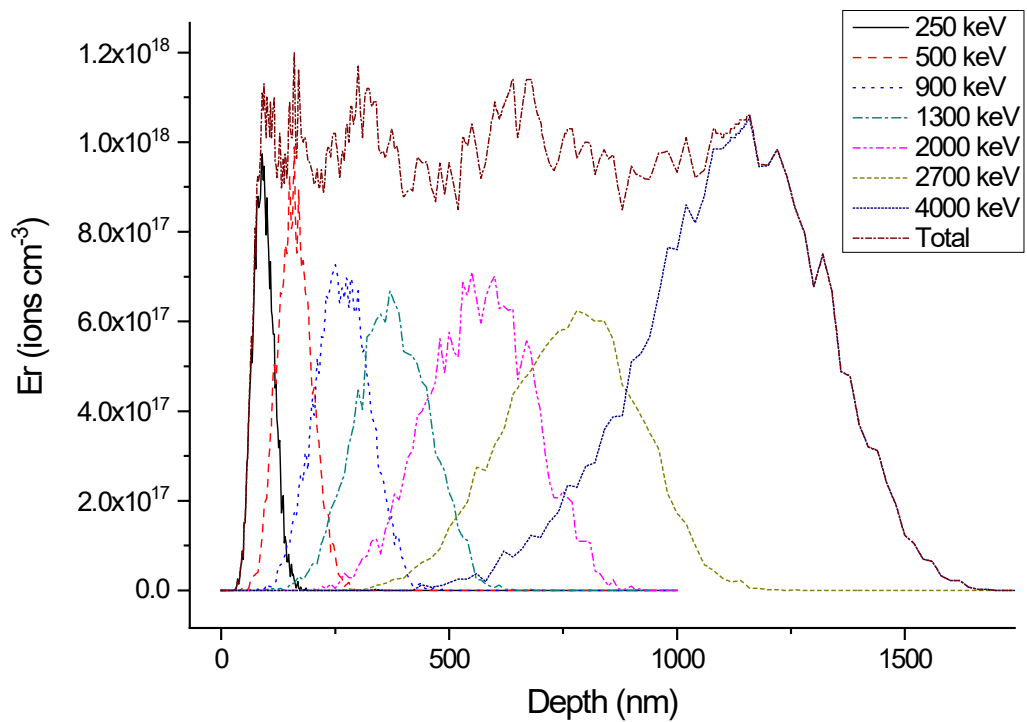

Supplementary Figure S1 Simulated implant profile, using SRIM, for Er with a peak concentration of $10^{18}$ cm$^{-3}$ and a total areal dose of $2.6\times10^{14}$ cm$^{-2}$

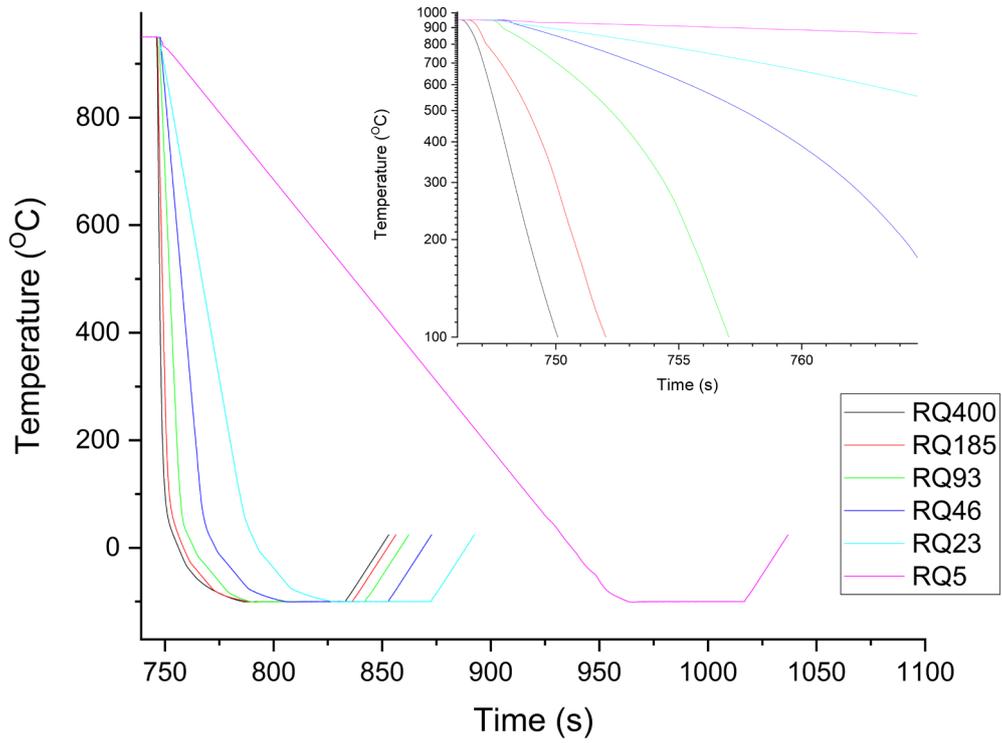

Supplementary Figure S2 Measured temperature decay profile in the quench phase of samples RQ400 to RQ5.

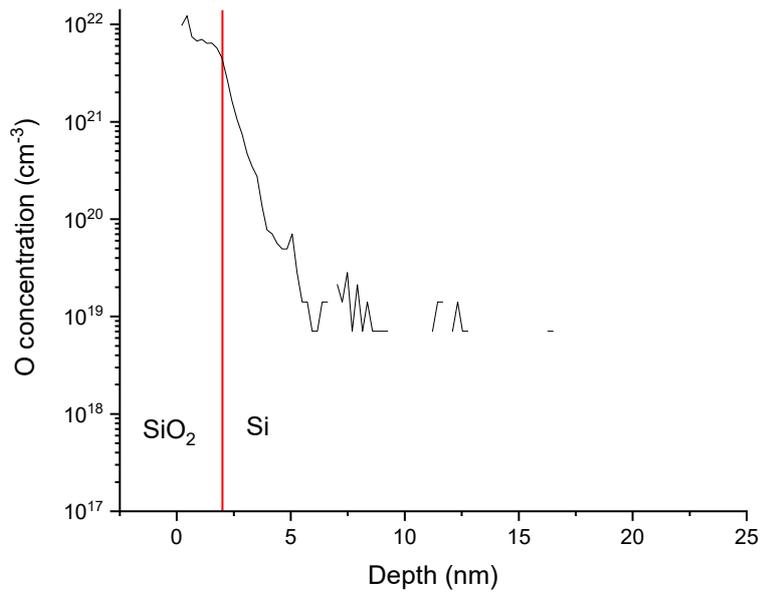

Supplementary Figure S3 Simulated O recoil implantation from 2 nm native oxide into Si from a weighted average of our Er chain implant with a total dose of $2.6\times10^{14}$ cm$^{-2}$.

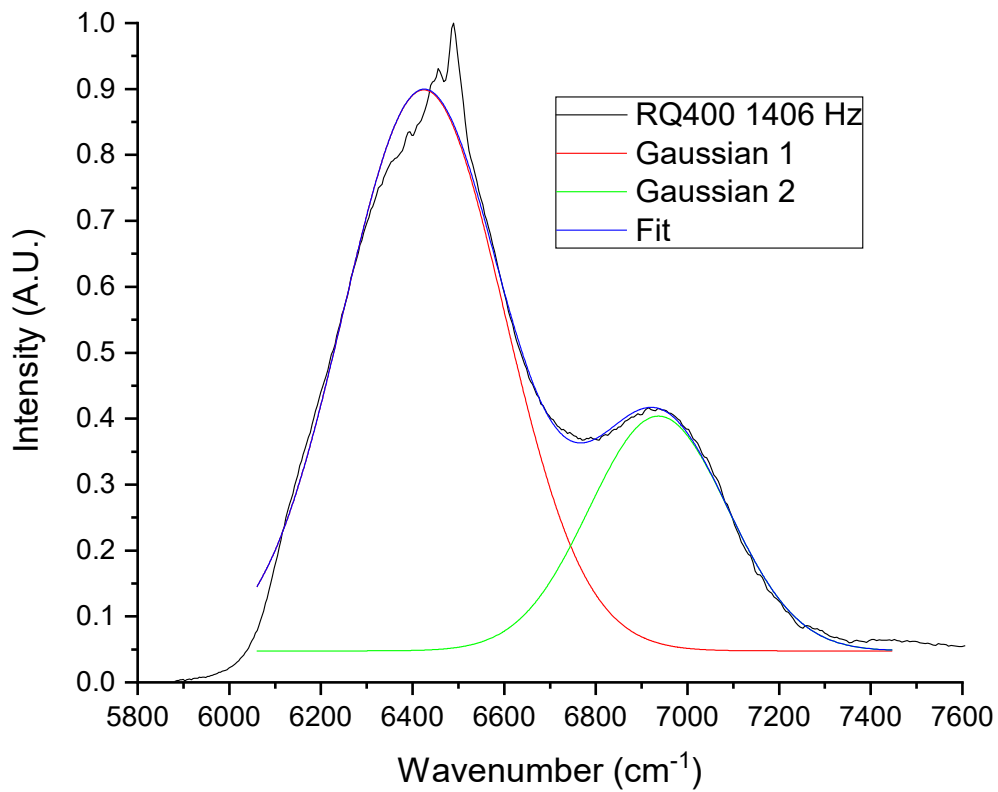

Supplementary Figure S4 PL from RQ400 at 3.5 K fitted with two Gaussians. The peak and FWHM are: Gaussian 1, 6425 cm$^{-1}$, 350 cm$^{-1}$; Gaussian 2, 6937 cm$^{-1}$, 303 cm$^{-1}$. In order to improve fitting to the broad Gaussian bands, we increased the modulation frequency to 1406 Hz which suppressed the longer lifetime Er PL peaks relative to the shorter lifetime broad Gaussian bands.

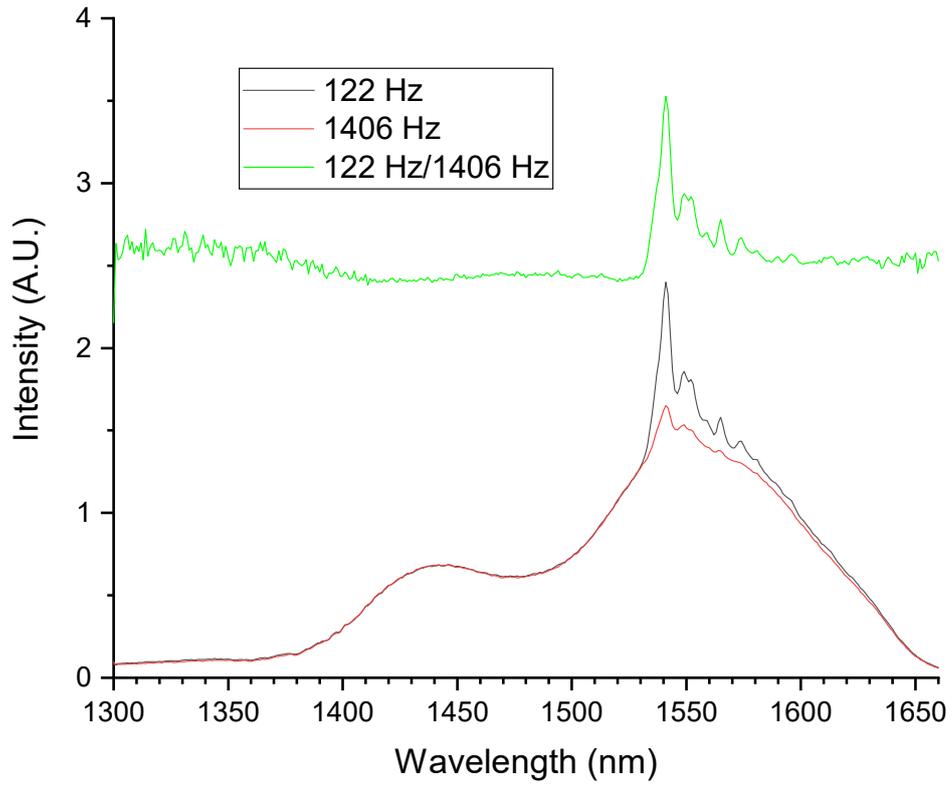

Supplementary Figure S5 PL spectra of RQ400 at 3.5 K taken with modulation frequencies of 122 and 1406 Hz, along with the ratio of these modulation frequencies. The 1406 Hz spectrum was normalised to show the overlap of the broad bands.

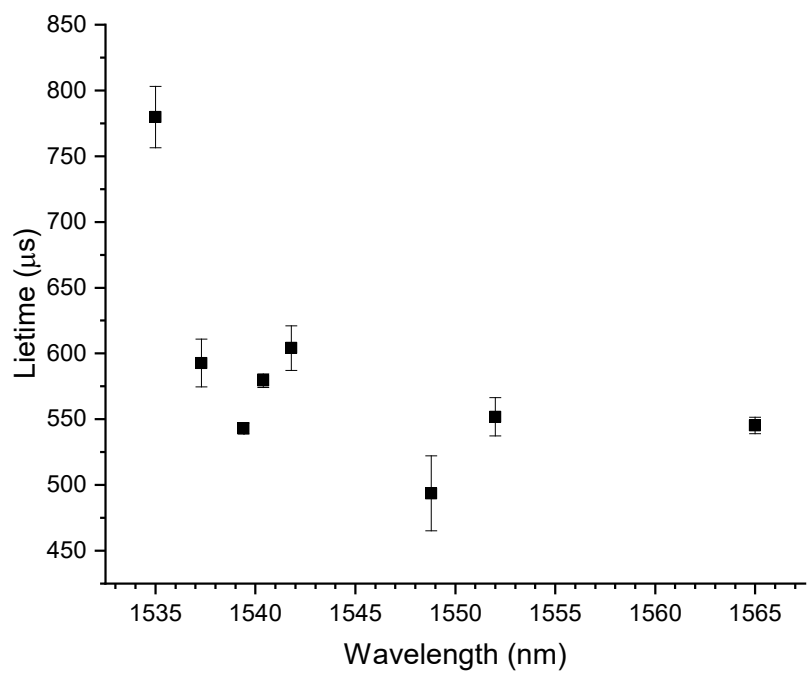

Supplementary Figure S6 Lifetime of the long lifetime component of RQ400 at 3.5 K. The resolution was 0.6 nm. In this measurement, the shortest lifetime component could not be resolved so a double exponential was sufficient to fit the decay. Only the longest lifetime component is of interest here. Ten separate decays for each wavelength were fitted with a double exponential and the error is the standard error of the mean.

Supplementary Table S1 $C_{2v}$ Orthorhombic CFPs (cm$^{-1}$) from fits to observed energy levels for various Er centres: Gritsch A, Gritsch B, Er-C, RQ-C, RQ-E

|  | Gritsch A | Gritsch B | Er-C | RQ-C | RQ-E |
|---|---|---|---|---|---|
| $B_0^2$ | -503.2 | -483.8 | -601.4 | -645.3 | -389.9 |
| $B_0^4$ | -1175.9 | -992.0 | -747.5 | -681.3 | -623.0 |
| $B_0^6$ | 113.5 | 69.5 | 202.2 | 111.7 | 180.7 |
| $B_2^2$ | 359.9 | 375.7 | 330.4 | -449.9 | -283.5 |
| $B_2^4$ | -100.5 | -160.7 | -199.7 | 62.5 | 137.0 |
| $B_4^4$ | 156.1 | 284.5 | 416.3 | 527.2 | 476.8 |
| $B_2^6$ | -371.4 | -306.6 | -236.1 | -46.2 | 59.5 |
| $B_4^6$ | 68.6 | 63.6 | 6.0 | -292.4 | -22.9 |
| $B_6^6$ | 362.3 | 417.9 | 473.4 | -383.9 | -322.7 |